\documentclass[reprint,amsmath,amssymb,aps,prl,groupedaddress,nofootinbib,twocolumn,superscriptaddress]{revtex4-1}
\usepackage{graphicx}
\usepackage{amsthm,amssymb,amsmath,braket,mathdots}
\usepackage{bm}
\usepackage[pagebackref=false,pdfnewwindow=true]{hyperref} 
\usepackage{epstopdf,psfrag}
\usepackage{relsize,amsbsy}
\usepackage[export]{adjustbox}
\usepackage{soul}
\usepackage{graphicx,xcolor,tikz}
\usepackage{float}

\DeclareMathOperator{\sgn}{sgn}

\newcommand{\be}{\begin{equation}}
\newcommand{\ee}{\end{equation}}

\newcommand{\bit}{\begin{enumerate}}
	\newcommand{\eit}{\end{enumerate}}

\definecolor{bananayellow}{rgb}{1.0, 0.88, 0.21}
\definecolor{straw}{rgb}{0.32, 0.28, 0.1}

\begin{document}
	\title{Floquet Time Spirals  and Stable Discrete Time Quasi-Crystals \\ in Quasi-Periodically Driven Quantum Many-Body Systems}
	\author{Hongzheng Zhao}
	\email{hongzheng.zhao17@imperial.ac.uk}
	\author{Florian Mintert}
	\author{Johannes Knolle}
	\affiliation{\small Blackett Laboratory, Imperial College London, London SW7 2AZ, United Kingdom}
	\date{\today}

	\begin{abstract}
	
	We analyse quasi-periodically driven quantum systems that can be mapped exactly to periodically driven ones
and find {\it Floquet Time Spirals} in analogy with spatially incommensurate spiral magnetic states.
Generalising the mechanism to many-body systems we discover  that a form of discrete time-translation symmetry breaking can also occur in quasi-periodically driven systems. We construct a discrete time quasi-crystal stabilised by many-body localisation.  Crucially, it persists also under perturbations that break the equivalence with periodic systems. As such it provides evidence of a stable quasi-periodically driven many-body quantum system which does not heat up to the featureless infinite temperature state. 
		
		
	\end{abstract}
	
	\maketitle
	\maketitle
	The dynamics of a quantum system governed by a generic time-dependent Hamiltonian is generally hard to analyse because to reach a certain point in time the entire history of evolution is required. Hence, simulations quickly become computationally unfeasible, especially for many-body systems. There are some exceptions such as periodically driven (PD) systems, i.e. whose Hamiltonian is periodic in time $H(t+T)=H(t)$, for which Floquet theory significantly reduces the effort to simulate the dynamics by decomposing the wave function to Floquet modes periodic over one time period $T$ up to phases. PD systems are of immense recent interest as they have been exploited extensively for engineering sought after Hamiltonians~\cite{Oka2009,Kitagawa2010,Lindner2011,Goldman2014,Cooper2019}, controlling quantum systems~\cite{DC,Flo2,Flo3,TC,SD,Colloquium,FE,BEC,Takashi2019}and realising non-equilibrium quantum phases~\cite{DPT,PSDQS,ESG}. For instance, the proper combination of periodic driving and many-body localisaton~\cite{Nandkishore2015} gave birth to the discovery of discrete time crystals (DTCs)~\cite{CFTC,DTC,FTC,TCPlatform,ObservationTC1,ObservationTC2,ObservationTC3,Sacha2017}, a new quantum phase with broken discrete time translation symmetry.
	
	Here, we address the intriguing question whether quasi-periodically driven (QPD) quantum systems can show features similar to time-crystalline behaviour of periodically driven ones? Normally it is expected -- apart for special integrable systems with momentum conservation~\cite{Maity2019} -- that QPD systems generically heat up to featureless infinite-temperature states~\cite{PotterAndrew2018,Rehn2016}. Hence, a related important question we address is whether there are interacting QPD  quantum systems with a stable non-trivial  steady state? Of course, generally such questions are hard to answer because Floquet theory, which provides an elegant framework for the analysis of PD quantum systems normally does not generalise to aperiodic time-dependencies. Here, we address the above questions by studying a class of QPD systems which can be efficiently treated within Floquet theory. This allows us to find QPD steady states with time-crystalline behaviour at special fine-tuned points whose stability to generic perturbations is studied with exact diagonalisation.
	
	\begin{figure}
		\centering
		\includegraphics[width=0.85\linewidth]{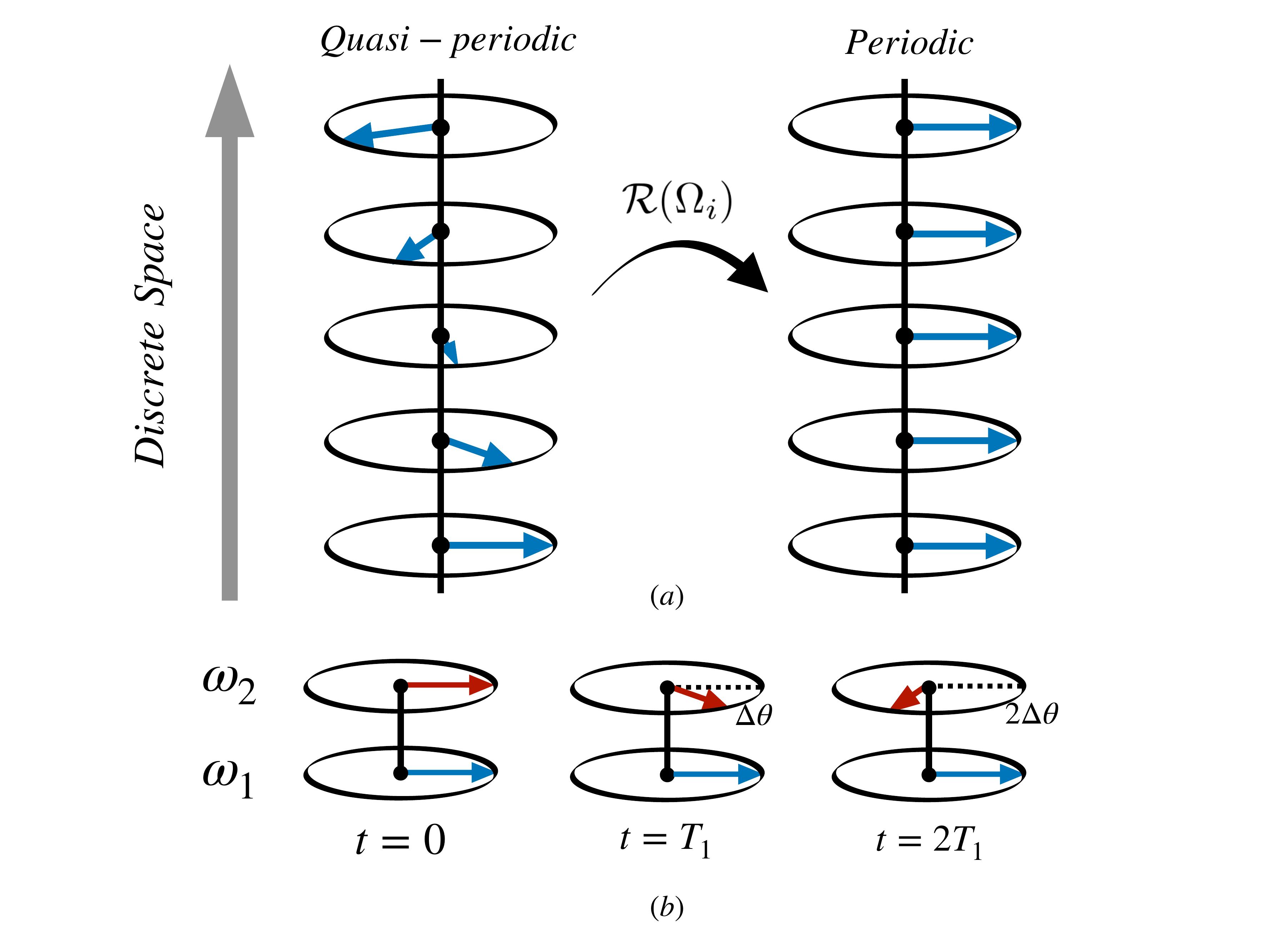}
		\caption{(a) Mapping a quasi-periodic magnetic spiral to a periodic ferromagnet. (b)Illustration of the two components of the QPD Hamiltonian on the Bloch sphere. The vectors represent the  driving fields in opposite directions with distinct frequencies $\omega_{1/2}$ . The blue comes back to the original configuration at multiple periods $nT_1$, while the red one with frequency $\omega_2$ never comes back with mismatched angle $n\Delta \theta$. }
		\label{fig:discrete}
	\end{figure}

	The main idea is that special cases of originally QPD Hamiltonian $H$ can be mapped exactly to a periodic counterpart $\tilde H$ in a rotated frame as
\begin{eqnarray}	
\label{eq.transformation}
\tilde{H}(t)=UH(t)U^{\dagger}-iU\partial_t U^{\dagger} ,
\end{eqnarray} 
 via time-dependent unitary transformations $U(t)$.
After solving the dynamics governed by the periodic Hamiltonian, the time-evolution in the original frame can be simply mapped back by applying the inverse of the unitary. The underlying idea -- albeit with {\it site}-dependent instead of {\it time}-dependent unitaries -- has been employed to study spiral magnetism in doped Hubbard models in the context of high-temperature cuprate superconductors~\cite{Schulz1990,Dopping,Kampf1996} and with recent cold atom experiments~\cite{DoppingExp}. For co-planar spirals the Hubbard model in the spin rotated frame becomes translationally invariant efficiently treatable via Bloch waves even for incommensurate spirals. Translating these ideas to the time domain of QPD systems allows us to construct {\it Floquet time spirals} (FTSs).

For pedagogical reasons, we provide an elementary example of a FTS via a two level system driven by two incommensurate frequencies which can be mapped to a periodic system. Generalising the idea to many-body systems we argue that a form of discrete time-translation symmetry breaking can also occur in QPD systems and 
we construct a minimal model of a discrete time quasi-crystal (DTQC). Crucially, we show that its appearance is not necessarily limited to the situations that are equivalent to a Floquet-problem, but we demonstrate that systems with many-body localisation show the DTQC behaviour even if perturbations break this equivalence.
As such, the exact solvability of the models derived with our present construction allows us to gain clear physical insight into phenomena that extends far beyond these special cases.

We are interested in QPD systems which can be defined via a Fourier expansion of the Hamiltonian~\cite{Crowley2019,TFC,QPFlorian}
	\begin{eqnarray}
	\label{eq.quasi}
	H(t) = \sum_{\vec{n}}H_{\vec{n}}e^{i\vec{n} \cdot \vec{\theta_t}},
	\end{eqnarray}
with $\vec{n} = (n_1,n_2,n_3,
	\dots)\subseteq \mathbf{Z}^N$ and $\theta_{t,i}=\omega_i t$, where all or a subset of the $N$ frequencies $\omega_i$ are incommensurate, i.e. their mutual ratios are irrational. There are some approaches to analyse QPD systems by mappings to higher dimensions~\cite{TFC,TClass}, constructing effective Hamiltonians to describe the system perturbatively~\cite{QPFlorian}, for analysing their asymptotic behaviour~\cite{Asymptotic,Nandy2017,Nandy2018} or using them for enhanced performance of quantum simulations~\cite{Ratchets,Controltransport} and creation of Majorana edge modes~\cite{Peng2018}. While in general for QPD systems we are lacking efficient methods for calculating their full time evolution, the FTSs introduced here are a special class corresponding to periodic systems in a rotating frame. 	
	
\textit{Mapping Quasi-Periodic to Periodic Systems.}
The basic idea goes back to studies of magnetic phases in the Hubbard model, where incommensurate spiral magnetic states can be induced by doping~\cite{Schulz1990,Dopping,Kampf1996}. There, spatially quasi-periodic spin spirals can be mapped to periodic ferromagnetic states by introducing site-dependent unitary transformation~\cite{Schulz1990,Dopping,Kampf1996}, which locally rotate the spin quantisation axis, see Fig.~\ref{fig:discrete}(a). Crucially, for co-planar spin spirals the rotated  Hubbard Hamiltonian is translationally invariant and can be conveniently described with momentum quantum numbers. 
The same idea can be extended to the time domain: We can construct instances of Hamiltonians $H(t)$, which are {\it aperiodic} in time such that $H(t)\neq H(t+T)$ for any $T$, but the time evolution of states $|\tilde{\phi}(t)\rangle = U(t)|{\phi}(t)\rangle $, which are rotated to a new frame via a time-dependent unitary transformation $U(t)$, is governed by a time $periodic$ Hamiltonian with $\tilde H(t)=\tilde H(t+T)$. Here $|\phi(t)\rangle$ is the time dependent solution of the Schr\"{o}dinger equation of the original system and the Hamiltonian in the rotated frame is given by Eq.~\ref{eq.transformation}~\cite{BEC}. 

In practice the construction is done in reverse such that any periodically time-dependent Hamiltonian helps to define an aperiodic Hamiltonian
	in terms of a time-dependent unitary transformation which allows us to analyse specific instances of quasi-periodic systems in terms of Floquet theory,
	and to understand their dynamics. Crucially, we show that some of the lessons learned from these specific instances directly translate to the general case where a mapping between periodic and quasi-periodic systems is missing. As such, we can use Floquet theory in order to understand the physics of systems to which Floquet theory fundamentally does not apply.
	
\begin{figure}
		\centering
		\includegraphics[width=0.8\linewidth]{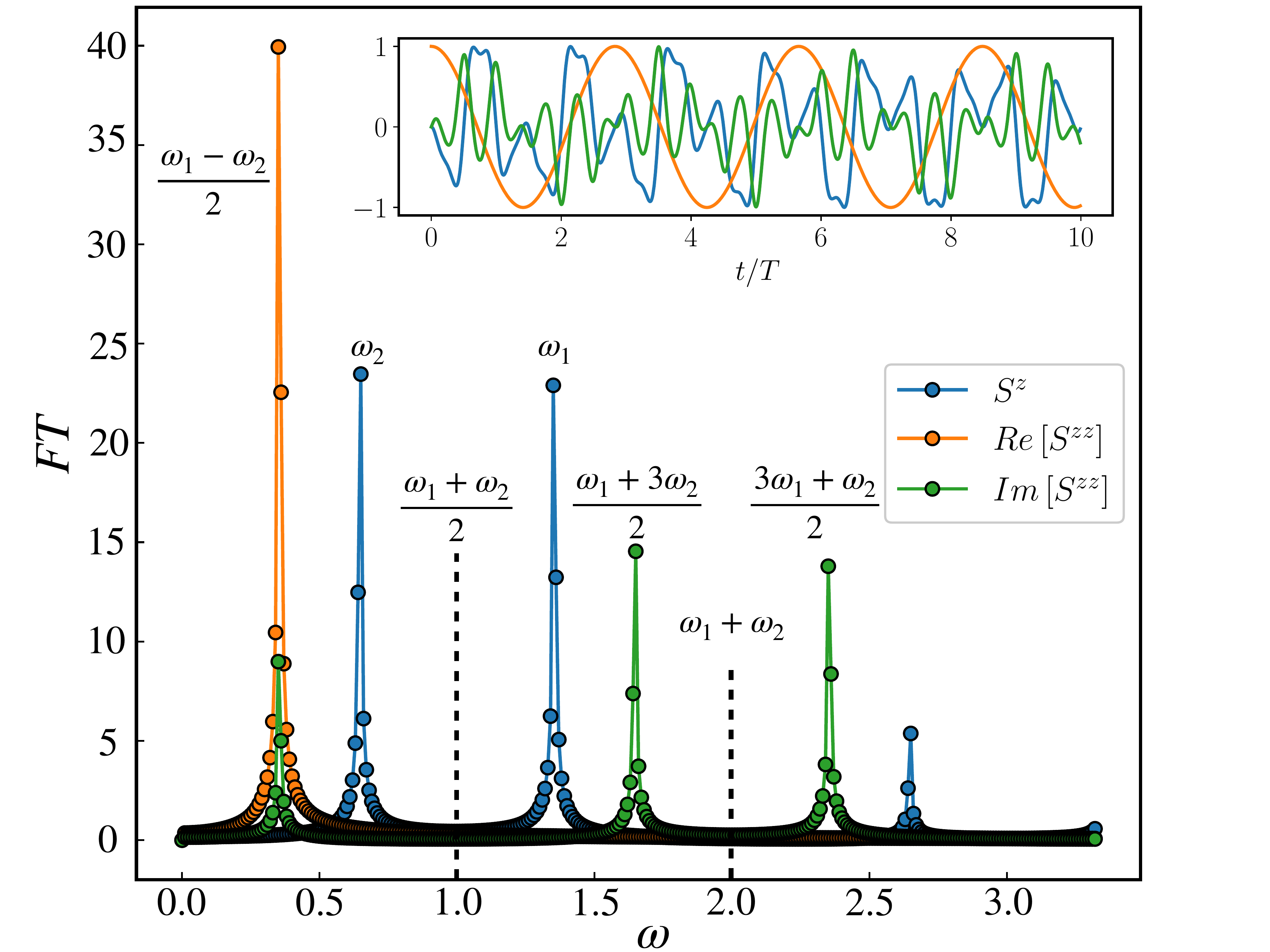}
		\caption{QPD two-level system: Fourier transformation of the dynamics of observables computed by exact diagonalization for parameters $A = 1, \omega = 1,\Omega=\sqrt{2}/4$. Peaks of $S^z$ appearing at $k \omega\pm\Omega$ for odd integer $k$ suggests that the system synchronises with the external drive. Dashed lines labels the frequencies of observables in the rotating frame. The inset presents real time behaviour.}
		\label{fig:fft_single}
	\end{figure}

	\textit{Floquet Time Spirals of a QPD Spin.} 
As an elementary example of a FTS we study a QPD two-level system with two frequencies $\omega_1$, $\omega_2$. 
	Consider the time-dependent Hamiltonian, $H(t) = H_1(t)+H_2(t)$, where $H_1, H_2$ describes two circular drivings
	\begin{eqnarray}
	\label{eq.time_spiral}
	\begin{aligned}
	{H}_1(t) &= -\frac{A}{2}\cos(\omega_1 t)\sigma_z+\frac{A}{2}\sin(\omega_1 t)\sigma_y,\\
	{H}_2(t) &= -\frac{A}{2}\cos(\omega_2 t)\sigma_z-\frac{A}{2}\sin(\omega_2 t)\sigma_y+\frac{\Omega}{2}\sigma_x,
	\end{aligned}
	\end{eqnarray}
	with Pauli matrices $\sigma_{\alpha}$ for $\alpha = x,y,z$ and the last term of strength $\Omega/2$ is a constant field in $x$ direction. One can plot the two components of the time-dependent Hamiltonian on the Bloch sphere as conversely rotating vectors illustrated intuitively in Fig.\ref{fig:discrete}(b).  We choose the driving component with $\omega_{1}$ to set a discrete time coordinate (blue) which now plays the role of a lattice.  Suppose the two frequencies are mutually incommensurate, the other rotating red vector with frequency $\omega_2$ in Fig.~\ref{fig:discrete} is aperiodic with a mismatched phase angle $n\Delta \theta = n\frac{\omega_2}{\omega_{1}}2\pi$ at discrete times $nT_1$. According to the definition given in Eq.~\ref{eq.quasi}, the Hamiltonian is certainly quasi-periodic. 
	Now the unitary transformation 
		\begin{equation}
U(t) = e^{i(\omega_1-\omega_2)t\sigma_x/4},
	\end{equation}
implements a rotation around the $x$-axis linearly changing in time and gives according to Eq.~\ref{eq.transformation}, the rotated Hamiltonian 
	\begin{equation}
	\tilde{H}(t) = -A\cos(\omega t)\sigma_z+(\frac{\Omega}{2}-\frac{\omega_{1}-\omega_2}{4})\sigma_x.
	\end{equation}
	It is simply a two-level system driven by a periodic field with frequency $\omega=(\omega_1+\omega_2)/2$, which has been studied with Floquet theory extensively~\cite{exact,Floexa}.  
 For the sake of simplicity we have set $\Omega = (\omega_1-\omega_2)/2$ such that the constant field vanishes and Floquet quasi-energies are zero, which permits an exact solution of the time evolution with
	\begin{equation}
	\label{eq.1BZ}
	|\tilde{\phi}(t)\rangle =  \sum_{\sigma}c_{\sigma}(t=0) \exp\Big[i   \text{sgn}[\sigma]\frac{A}{\omega}\sin(\omega t)\Big]|\sigma\rangle,
	\end{equation}
where $\sigma = \uparrow,\downarrow$ and $\sgn[\sigma ]$ gives $+,-$ respectively.
	For an initial state with $c_{\sigma}(t=0)=1/\sqrt{2}$ one readily obtains the magnetisation in the $z$ direction 
	\begin{eqnarray}
	\label{eq.S_Q}
	\begin{aligned}
	S^{z}(t) & =  \langle {\phi}(t)| \sigma^{z}|{\phi}(t)\rangle= \langle \tilde{\phi}(t) |U(t)\sigma^{z}U^{\dagger}(t)|\tilde{\phi}(t)\rangle\\
	&= -\sin(\Omega t)\sin\left[\frac{2A}{\omega}\sin(\omega t)\right],
	\end{aligned}
	\end{eqnarray}
	where the time dependence contains two independent frequencies, as shown in the inset of Fig.~\ref{fig:fft_single}. 
Other observables, like non-equal time correlation function $S^{zz}(t,0)=\langle {\phi}| \sigma^{z}(t)\sigma^{z}(0)|{\phi}\rangle =\cos(\Omega t) + i\sin(\Omega t)\cos\left[\frac{2A}{\omega}\sin(\omega t)\right]$ can be computed in a similar manner.
	Note, the imaginary part of this correlation function can be treated as a suitable measure for the quasi-periodicity.
The dominating peaks in the frequencies spectrum of the magnetisation $S^z$, blue line of Fig.~\ref{fig:fft_single},  are located at $\frac{k\pm1}{2}\omega_1+\frac{k\mp1}{2}\omega_2$ or equivalently   $k\omega\pm\Omega$  for odd integer $k$, ensuring that only integer multiples of $\omega_{1/2}$ appear like the first two blue peaks located at $\omega_{1},\omega_2$. This shows that the observable simply synchronises with the external drive as expected. The fact that the peaks of the Fourier decomposition of the Hamiltonian coincide with the one of the observables is a signature that the system preserves the long-range quasi-periodic time order imposed by the external drive, which is analogous to observables of well known spatial quasi-crystals which do not possess translational symmetry but are still long-range ordered~\cite{Senechal1996}. However, an intriguing possibility would be the breaking of this QPD long-range order, i.e. Fourier peaks at half integer multiples of the driving frequencies may appear as discussed in the following section.

\textit{Discrete Time Quasi-Crystals.}
The idea of FTS can be generalised to study QPD many-body systems like interacting spin chains. Here, we focus on a new type of DTQC and discuss its rigidity which shows that it is robust beyond our method of construction.

First, let us formally define what we mean by the breaking of long-range quasi-periodic time order. Recall the usual definition of discrete time translational symmetry breaking (TTSB)~\cite{FTC}. For a PD system with period $T$, there exists an operator $\hat{O}$ signalling TTSB if $\langle \hat{O}\rangle_t$ is only invariant under time translation by $nT$ for $n>1$~\cite{NoteTTSB}. In contrast to PD systems, QPD systems do not have a well defined period but, loosely speaking~\cite{NoteQTTSB}, one can still define quasi-periodic time translation symmetry breaking (QTTSB)  in terms of frequency. For instance, consider a QPD Hamiltonian defined by Eq.~\ref{eq.quasi}, QTTSB occurs if for each $t$,  there exists an operator $\langle \hat{O}\rangle_t$ with Fourier decomposition as
\begin{eqnarray}
\label{eq.QTTSB}
\langle \hat{O}\rangle_t = \sum_{\vec{n}}O_{\vec{n}}e^{i\sum_in_i\omega_i/p_it},
\end{eqnarray}
with at least one of the non-negative integers  $p_i$ larger than 1. Basic TTSB in PD system is included in the above definition, i.e. period doubling with only one $\omega_0$  and $p_0=2$. Note, there is no general concensus and our definition is different from a very recent proposal of time quasi-crystals in bosonic systems~\cite{giergiel2018} where the frequency of the drive and of the observable are not simply related, which is similar to other DTQCs studied before~\cite{Tongcang2012,Huang2018,Autti2018}.

Combining a standard DTC model~\cite{DTC} with the FTS construction allows us to write down a QPD model realising QTTSB. The Hamiltonian is given by
	\begin{eqnarray}
	\label{eq.DTC}
	H(t) = \label{key}H_{MBL}(t)+H_d(t)+\frac{\Omega}{2} \sum_j\sigma_j^x,
	\end{eqnarray}
	where 
	\begin{eqnarray}
	\label{eq.H_MBL}
	\begin{aligned}
	{H}_{MBL}(t) =h_1(t)\sum_{j=1}^{N}\left[ J_j\sigma_j^x\sigma_{j+1}^x-h_j^x\sigma_j^x\right],
	\end{aligned}
	\end{eqnarray}
	with $J_j,h_j^x$ chosen from uniform distributions. The precise role of disorder in nearest neighbour interactions and longitudinal fields in the formation of DTC is under debate~\cite{Sacha2015A,Ho2017A,Zeng2017A,Pal2017A,Rovny2018}. Here, disorder is applied to induce many-body localisation for generic small perturbations which prevents the system from heating to infinite temperature~\cite{PSDQS}.
	 Each spin is subjected to a QPD field captured by
	\begin{equation}
	\label{eq.H_d}
	\begin{aligned}
	{H}_d(t) = h_2(t)\sum_j\left[\cos(\Omega t)\sigma_j^z-\sin(\Omega t)\sigma_j^y\right].
	\end{aligned}
	\end{equation}
The driving terms in the Hamiltonian are switched on and off periodically such that the stepwise functions within one period $t\in (-T/2,T/2)$ read
	\begin{equation}
h_1(t)=
	\begin{cases}
	1, & |t|\leq \frac{T}{4} \\
	0, & |t|>\frac{T}{4}
	\end{cases},\ \ \ 
	h_2(t)=
	\begin{cases}
	0, & |t|\leq \frac{T}{4} \\
	g, & |t|>\frac{T}{4}
	\end{cases},
	\end{equation} 
ensuring that the time evolution governed by $H_d,H_{MBL}$ happens in two different time windows.
	The combined driving in Eq.~\ref{eq.H_d}, $	h_2(t)\cos(\Omega t) $ or $	h_2(t)\sin(\Omega t) $,  fits in the definition of quasi-periodicity Eq.~\ref{eq.quasi}
, which can be verified by using the Fourier expansion of the box function, e.g.
	\begin{eqnarray}
	\label{eq.quasibox}
	h_2(t)\cos(\Omega t) 
	= \sum_{n_1=-\infty}^{+\infty} \sum_{n_2=\pm 1} c_{n_1}e^{i(n_1\omega+n_2\Omega) t},
	\end{eqnarray}
	with $\omega=2\pi/T$ and coefficients $c_{n_1=0} = {g}/{2}$, and $-g\ \text{sinc}({n_1}/{2})/2$ for non-zero $n_1$. 
	According to Eq.\ref{eq.transformation} the unitary transformation $U(t) = e^{i\frac{\Omega}{2} t\sum_{j}^{N}\sigma_j^x}$
gives the new Hamiltonian 
	$\tilde{H}(t) = \tilde{H}_{MBL}(t)+\tilde{H}_d(t)$
	written as
\begin{equation}
	\label{eq.MBL}
	\tilde{H}(t) = h_1(t)\sum_j ^N\left[J_j\sigma_j^x \sigma_{j+1}^x-h_j^x\sigma_j^x\right] + h_2(t)\sum_j\sigma_j^z.
\end{equation}
	In the rotated frame it is the standard model of a DTC with well-defined period $T$ which for $g\approx\omega/2$ shows robust TTSB~\cite{FTC,DTC}. It is clear that TTSB in the rotating frame translates into QTTSB in the original physical frame.

\begin{figure}
		\centering
		\includegraphics[width=0.8\linewidth]{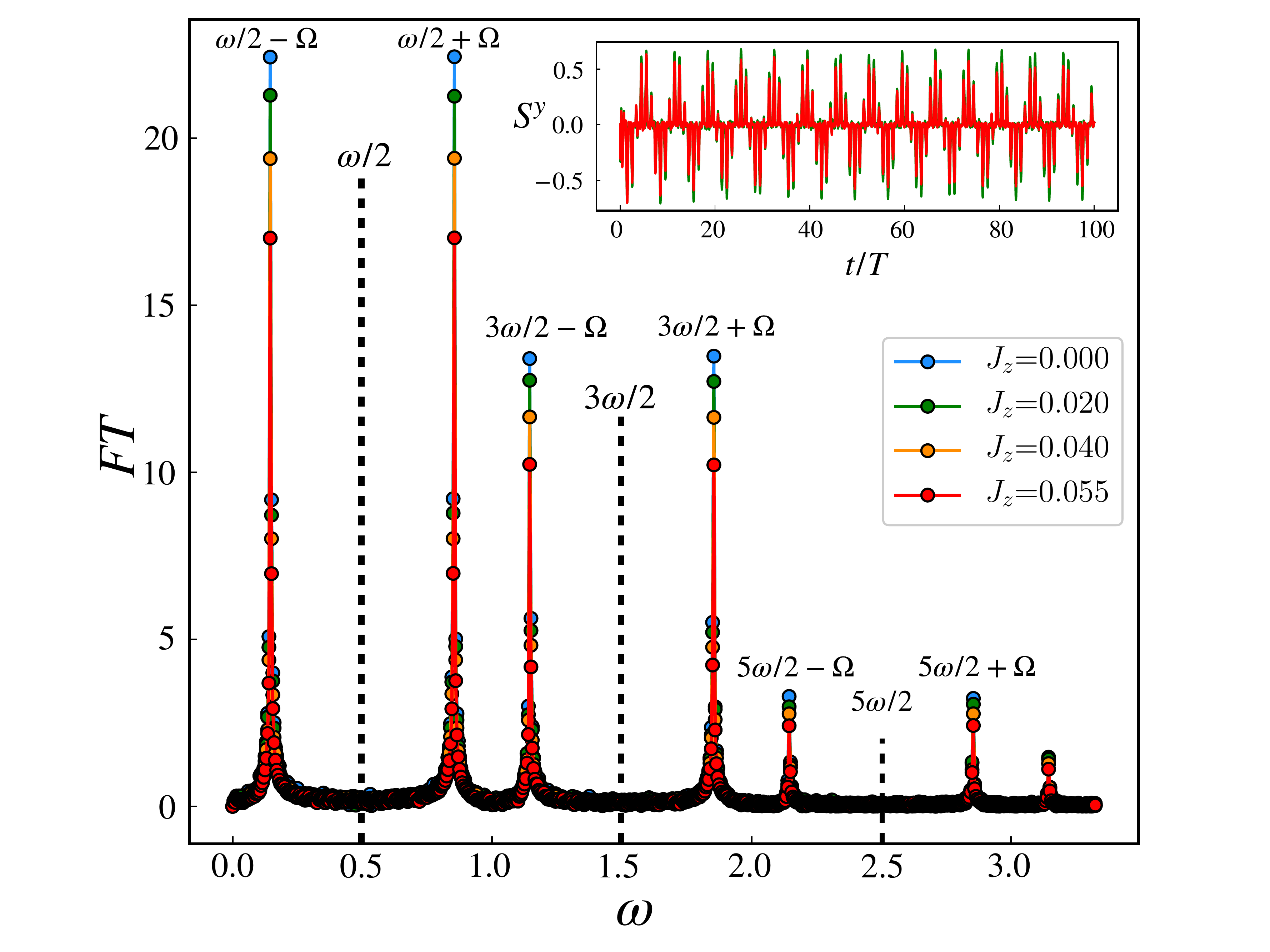}
		\caption{Fourier transformation (FT) of the average magnetization in y direction of the DTQC, computed by ED with 10 sites (average over 300 disorder configurations) for different interaction perturbations $J_z$. FT is done in a window 200-400 $T$.  For the system without perturbation, in the rotating frame, subharmonic behaviour can be identified by peaks at $k{\omega}/{2}$ with odd integer $k$ labelled by the dashed line, whereas in the original frame, new peaks at $k{\omega}/{2}\pm\Omega$,  show the existence of the robust time quasi-crystal.  Inset shows real time dynamics for $J_z=0.02, 0.055$.} 
		\label{fig.interperturb}
	\end{figure}
	
\textit{Signature of QTTSB, Rigidity and Thermalization.}  
QTTSB can be observed from the expectation values of the spin operators $S^{y,z}(t)$ ($\sigma^x$ commutes with the unitary transformation). In the following, we use ED to do numerical simulation given parameters $J_i\in [1/2,3/2],  h_j^x\in [0,1], \omega = 1,\Omega = \sqrt{2}/4$ with periodic boundary condition and trotter step $\Delta t =T/300$. Real time dynamics and the Fourier transformation of $S_y(t)$ are presented in Fig.~\ref{fig.interperturb}. The subharmonic response w.r.t. the Fourier decomposition of the drive in Eq.\ref{eq.quasibox} is a clear signature of QTTSB with sharp peaks located at $k\omega/2\pm \Omega$ for odd integer $k$.
	
	Finally, we check the stability of the DTQC by probing the rigidity of the $k\omega/2\pm\Omega$ subharmonic response in the Fourier spectrum. By construction, our system is stable against those perturbations, namely the pulse imperfections in $H_d$ or unwanted field disorders in $H_{MBL}$, which can be mapped to those studied before for the periodic system in the rotating frame~\cite{FTC,DTC}. 
Our major concern -- also to show that the DTQC is more fundamental than our construction of mapping it to a DTC --  are perturbations which can not be transformed to a simply periodic system in the rotating frame. The expectation is that generic QPD quantum many-body systems heat up to featureless infinite temperature states in the long time limit~\cite{PotterAndrew2018}. For concreteness we focus on adding interaction perturbation $h_1(t) \sum_j J_z\sigma_j^z \sigma_{j+1}^z$ to $H_{MBL}$ in Eq.~\ref{eq.H_MBL} of uniform strength $J_z$, which does not commute with the unitary transformation. Therefore even in the rotating frame, the system is QPD and probes the generic regime beyond our FTS construction which we analyse via exact diagonalization.

\begin{figure}[h]
	\centering
	\includegraphics[width=0.49\linewidth]{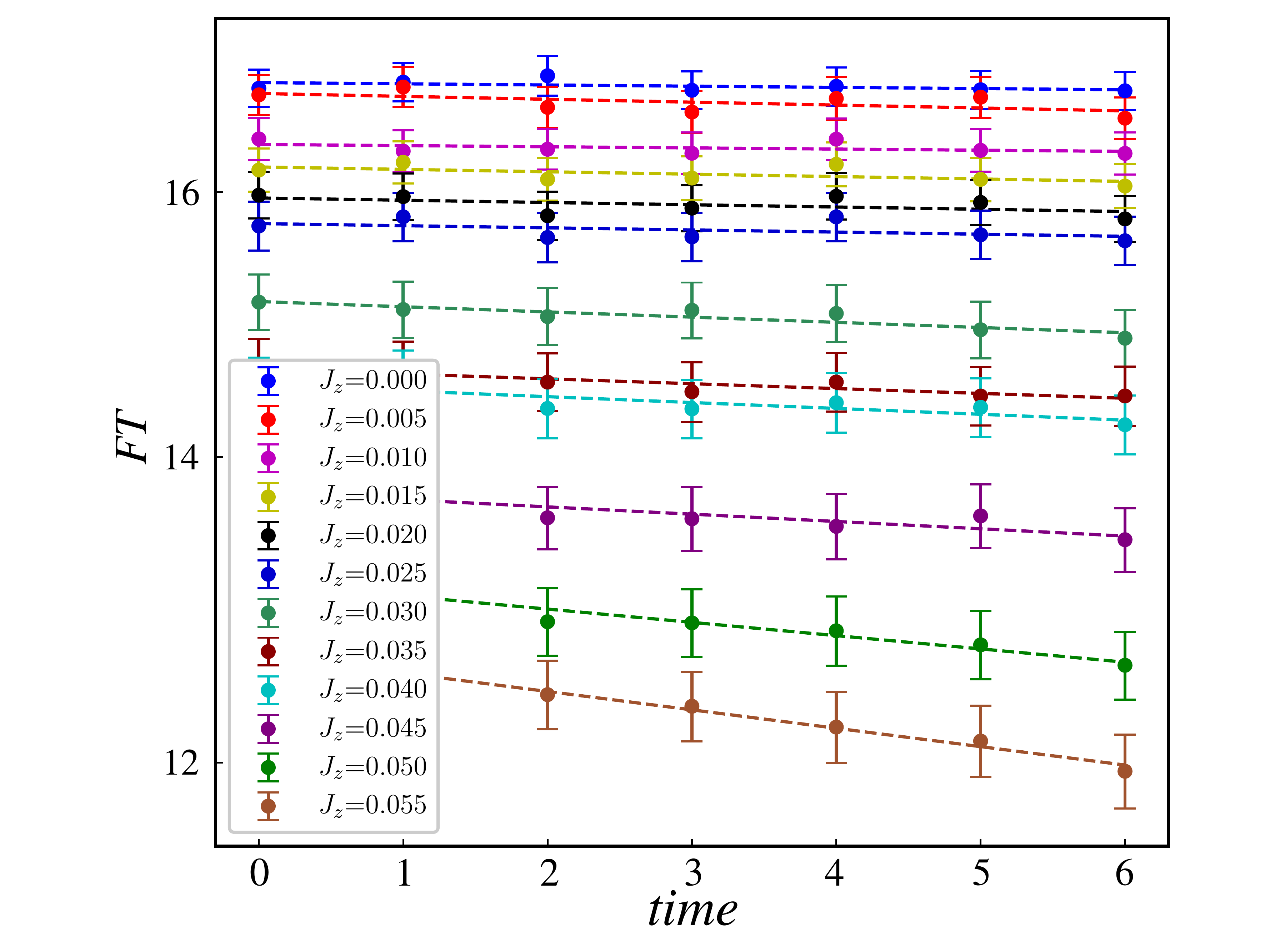}
		\includegraphics[width=0.495\linewidth]{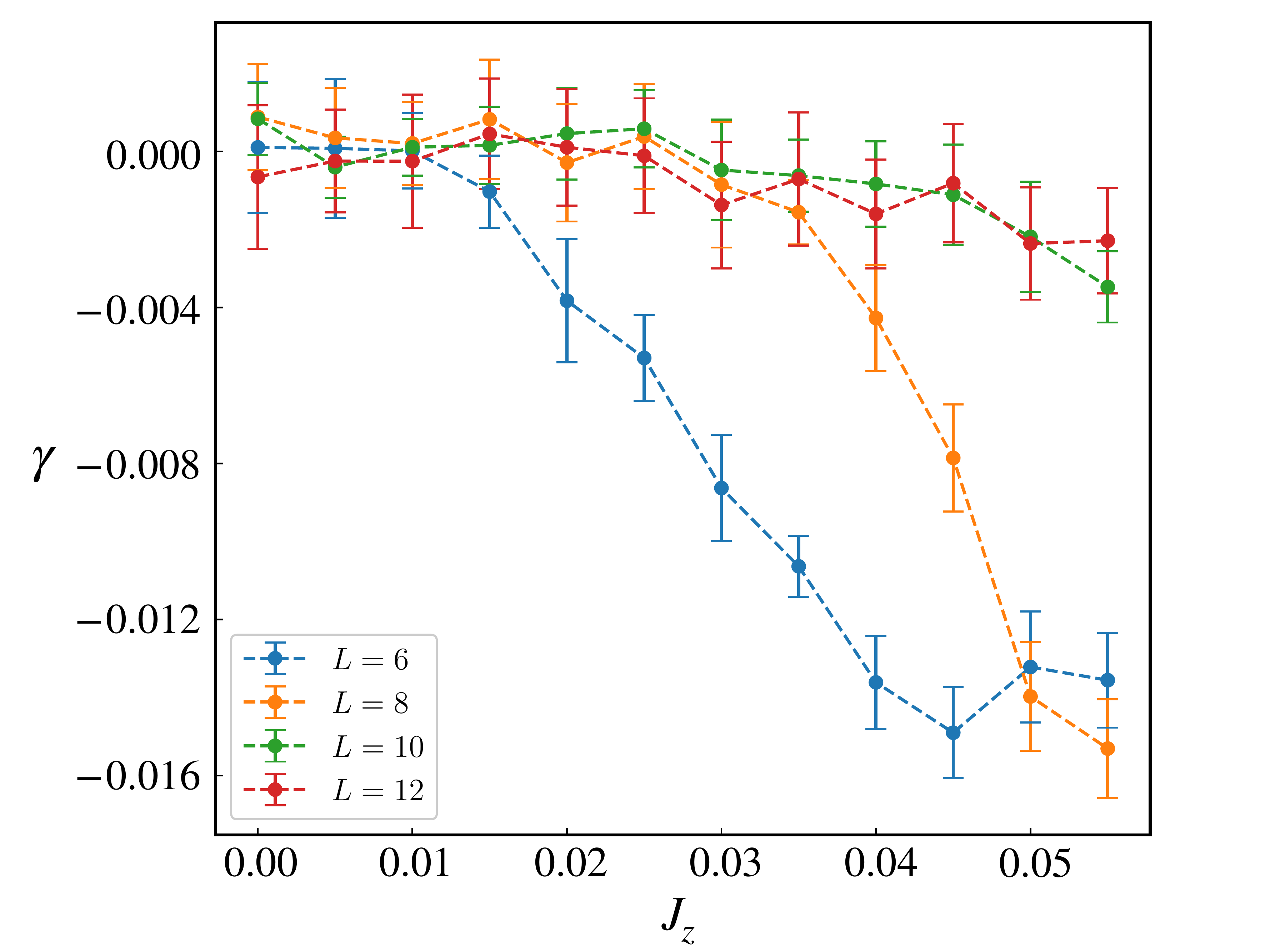}
	\caption{Left: Amplitudes of the first subharmonic peak of the Fourier spectrum of $S_y(t)$ for the DTQC on a log scale, as a function of increasing time windows (300 disorder configurations) for different interaction perturbations with $J_z$. The linear fit gives the decay rate $\gamma$. Right: $\gamma$ versus $J_z$ for varying system sizes.}
	\label{fig:fftpeak_main}
\end{figure}

The Fourier spectrum is shown in Fig.~\ref{fig.interperturb} for various values of $J_z$. Robust peaks are locked at  $k\omega/2\pm\Omega$ with descending heights for increasing perturbations. 
We check the long time stability of the DTQC phase by studying the amplitude of the first subharmonic peak as a function of shifting time windows with varying perturbations. The Fourier transformation is done within 7 equally divided time windows between the domain $[300T,1000T]$. Dotted lines are linear fits, of which the slopes $\gamma$ capture the rate  of decay of the signal due to thermalisation. We plot $\gamma$ w.r.t. $J_z$ for different system sizes to investigate the thermalisation transition induced by $J_z$. After a clear stable regime with zero decay rate, $\gamma$ drops suggesting the onset of heating. For instance, we find that for $J_z=0.02$, the oscillations persist without decay up to at least 1000$T$, demonstrating that the DTQC is robust w.r.t. small perturbative interaction. Note, the critical value of the phase transition is still size dependent but it increases to larger values with increasing system size. A precise determination of the transition point is beyond the scope of this work.   Additional simulations probing the stability with pulse imperfections are given in the Supplementary Material.

{\it Discussion and Conclusion.}
It is important to show that our DTQC is robust w.r.t. perturbations that do not permit a mapping to a periodic system, therefore, it is more fundamental than our fine-tuning FTS construction.  Further more, the numerical evidence for a stable QPD quantum many-body system is remarkable beyond the time crystal aspect. Normally, it is expected that QPD systems, e.g. driven by a Fibonacci sequence, thermalize even if the system is integrable~\cite{PotterAndrew2018} but the counter example presented here motivates more systematic studies of heating in QPD quantum systems.

	We have restricted our basic examples to unitaries of spin rotations varying linearly in time but it would be interesting to construct QPD systems with more general time- and space-dependent transformations. A more ambitious goal would be to start from generic QPD systems and investigate the existence of maps to periodic ones.
	The concept of DTQC substantially enriches our understanding of discrete time translation symmetry and its breaking, indicating that subhamonic frequency response is more fundamental than period n-tupling~\cite{Surace2019} ill-defined for QPD systems in real time. Although the QTTSB can be expected with subharmonic responses associated with all primary frequencies, e.g. $\omega$ and $\Omega$ in our case, we only observe the QTTSB with $\omega/2$ but not $\Omega$. It will be interesting to explore more general QTTSB and other DTQCs in the future.

	Finally, it will be worthwhile to explore the underlying idea of mapping quasi-periodic systems to periodic ones in other contexts beyond the original spiral magnets or the time domain as presented here. 
	
	{\it Acknowledgements.}
	We thank Achilleas Lazarides, Andrew Potter, Guanchen Peng and Shtitadi Roy for helpful discussions.
	This work is part of the project TheBlinQC that has received funding from the QuantERA ERA-NET Cofund in Quantum Technologies implemented within the European Union's Horizon 2020 Programme and from EPSRC under the grant EP/R044082/1.
	
	\bibliography{QP_ref}

\begin{thebibliography}{58}
\bibitem [{Not({\natexlab{a}})}]{NoteTTSB}%
  \BibitemOpen
  \href@noop {} {\  ({\natexlab{a}})},\ \bibinfo {note} {additional requirement
  should be satisfied as well, e.g. the state at time $t$ needs to be
  short-ranged correlated}\BibitemShut {NoStop}%
\bibitem [{Not({\natexlab{b}})}]{NoteQTTSB}%
  \BibitemOpen
  \href@noop {} {\  ({\natexlab{b}})},\ \bibinfo {note} {strictly speaking it
  is not the breaking of a symmetry but rather the long-range quasi-periodic
  order in time imprinted by the Hamiltonian drive. However, in analogy with
  the widely used TTSB we use the term QTTSB.}\BibitemShut {Stop}%
\end{thebibliography}

\clearpage
	\appendix
	\section{Supplementary Material
		\\Rigidity of Discrete Time Quasi-Crystal Phase}
	The quasi-energy level statistics has been employed to determine the thermalisation phase transition point of DTCs~\cite{DTC,PSDQS}. However, the perturbations we are concerned with here do not permit us to map it to a simple local PD Hamiltonian such that the Floquet operator and quasi-energies are not well-defined for the QPD system.
	Therefore, we study the perturbation's effects to DTQC in real time by ED up to 1000T, with trotter step $\Delta t =T/300$ with varying system size $L=6,8,10,12$, to analyse the thermalisation transition numerically.
\subsection{Interaction perturbation}	
\begin{figure}[h]
	\centering
	\includegraphics[width=0.8\linewidth]{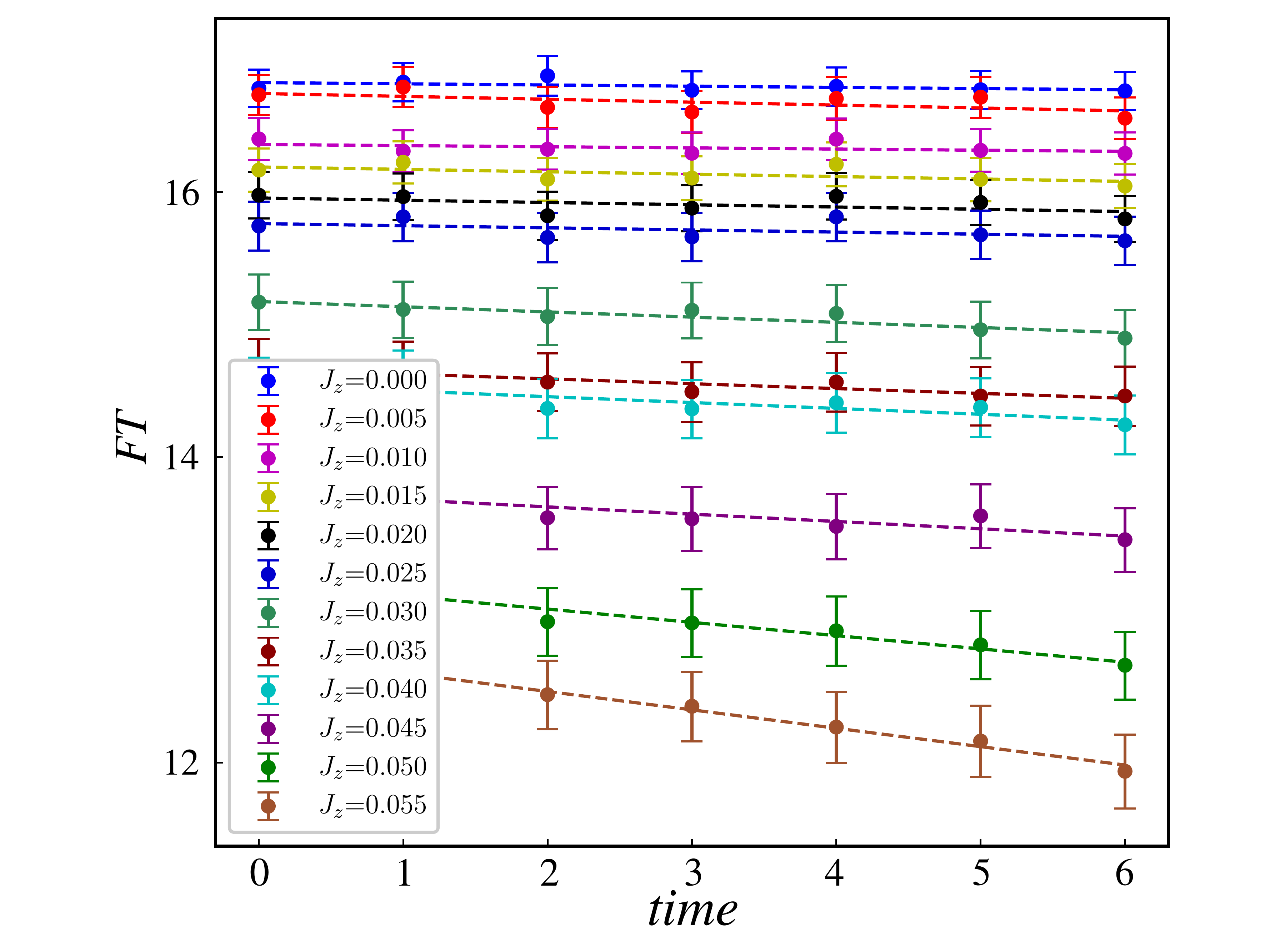}
	\caption{Amplitudes of the first subharmonic peak of the Fourier spectrum of $S_y$ for the DTQC versus time (300 disorder configurations) for different interaction perturbations.}
	\label{fig:fftpeak_int}
\end{figure}

We replot the Fig. \ref{fig:fftpeak_main} and explain details of numerical methods. 
In order to check the stability of the DTQC given perturbative interactions, in Fig.~\ref{fig:fftpeak_int}, we check the amplitudes of the first subharmonic peak in the Fourier spectrum for various perturbations $J_z$ as  a function of different shifting time windows computed with 10 sites.
FFT is done within 7 equally divided time windows between the domain $[300T,1000T]$. Error bars are depicted according to the standard deviation of mean, $\sigma/\sqrt{N}$ with $N$ being the number of disorder realizations. Dotted lines are linear fits, of which the slopes $\gamma$ capture the rate  of decay of the signal due to thermalisation. The existence of a clear regime with $\gamma \sim 0$ (horizontal line) is crucial, confirming that the new phase is robust and does not heat up beyond the fine-tuned point of construction; while for larger $J_z$, amplitudes of the subharmonic behaviour drop and the system thermalises. 

\begin{figure}[h]
	\centering
	\includegraphics[width=0.85\linewidth]{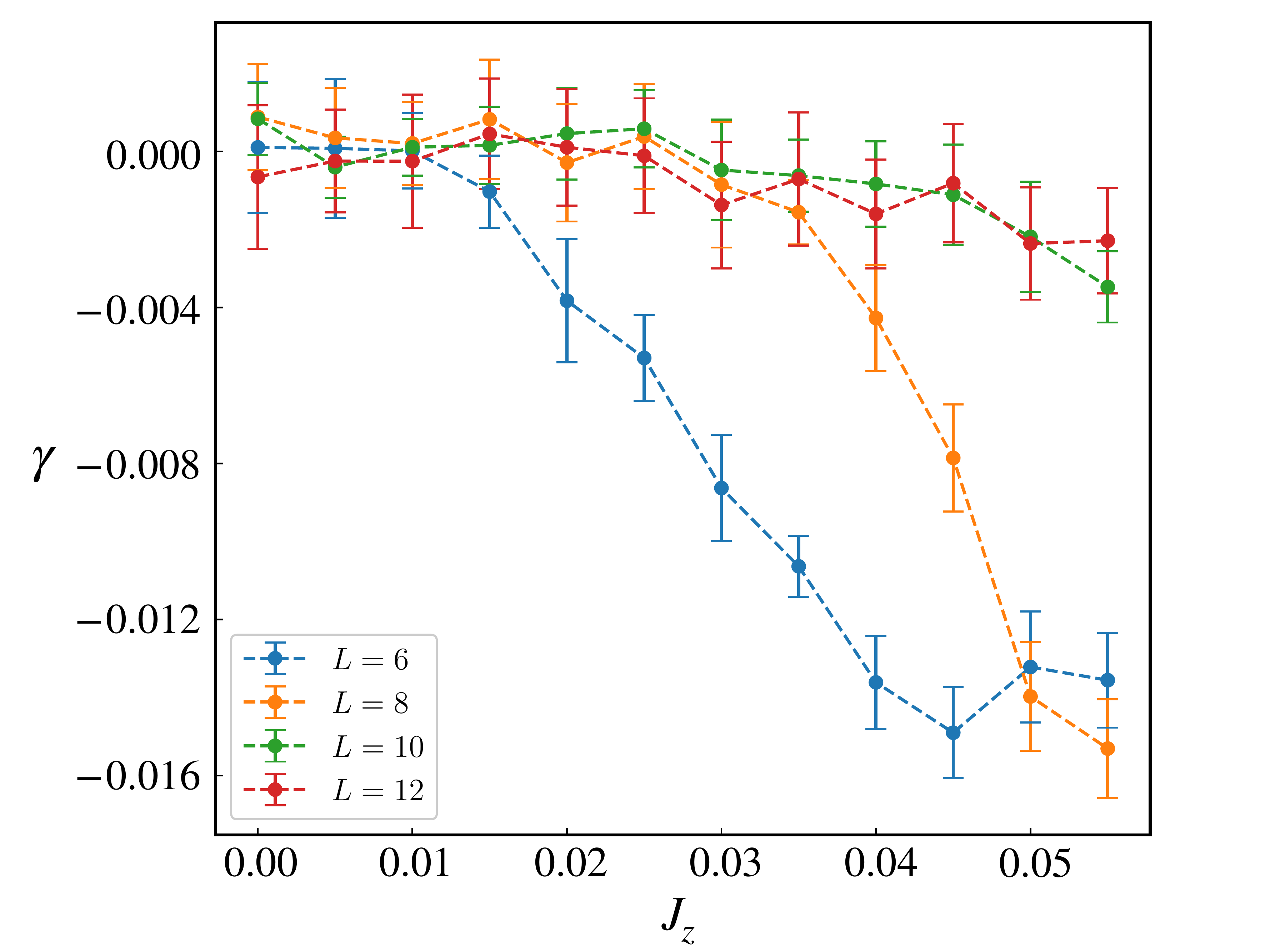}
	\caption{Decay rate $\gamma$ of the first subharmonic peak in Fourier spectrum versus perturbation $J_z$ for varying system sizes. Decay $\gamma$ remains zero for small perturbations and its drop indicates the transition to the thermalised phase. The transition point locates within $0.03-0.045$.}
	\label{fig:decay_int}
\end{figure}

In order to investigate the relation between decay, strength of perturbations, as well as finite size effects, we plot $\gamma$ as a function of $J_z$ for system sizes 6,8,10,12 in Fig.~\ref{fig:decay_int}. Each data point is obtained by averaging the $\gamma$ calculated with varying lengths of FFT time window from $70$ to $100T$. The standard deviation is presented as error bar.  After the stable regime with zero decaying rate, $\gamma$ drops suggesting the onset of thermalisation. The critical value of the phase transition varies and becomes larger with the system size. A precise determination of the transition point is beyond the scope of this work.

\begin{figure}[h]
	\centering
	\includegraphics[width=0.8\linewidth]{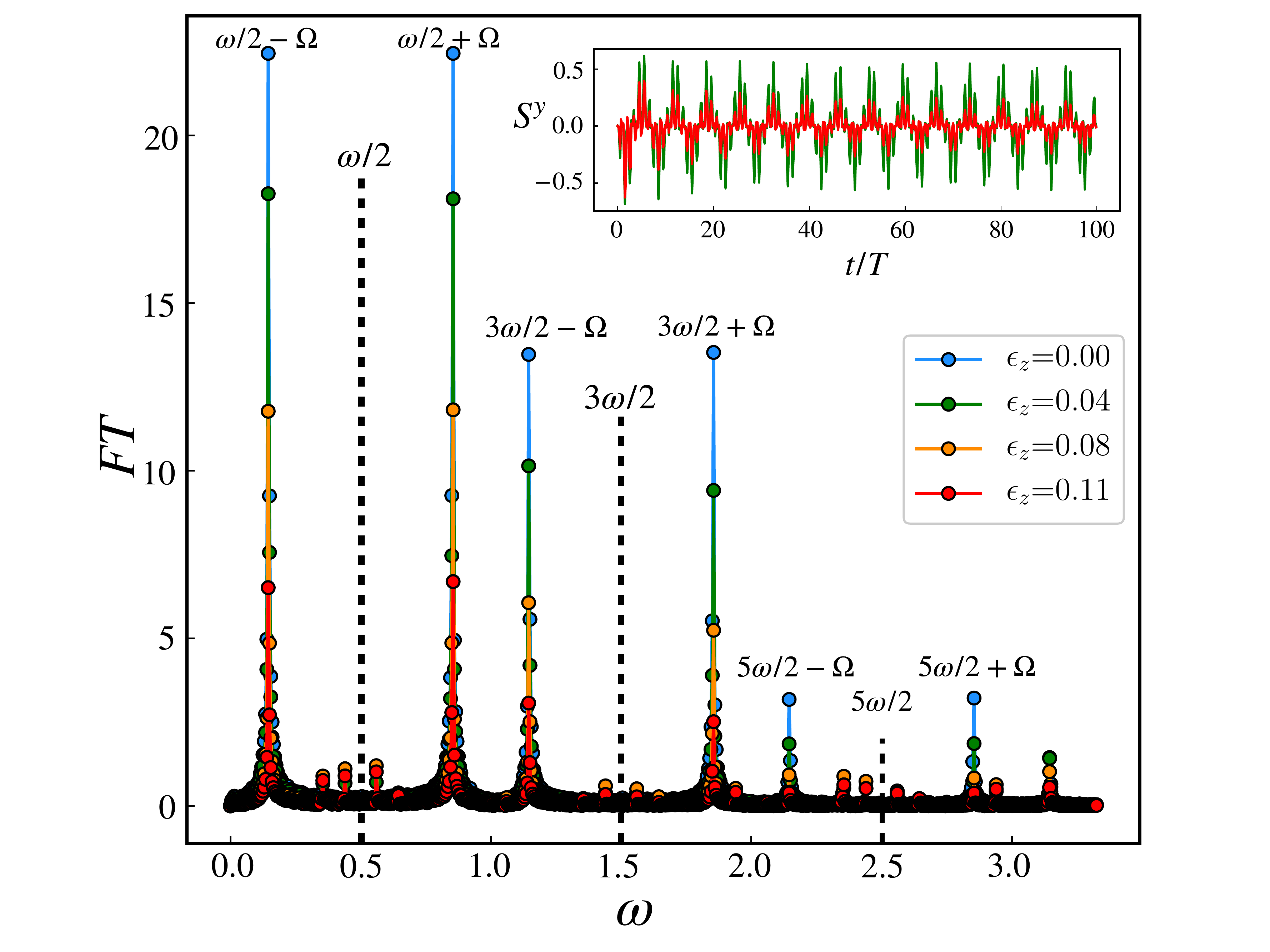}
	\caption{Fourier transformation of the average magnetization in y direction of DTQC, computed by  ED with 10 sites  (300 disorder configurations) for different $\epsilon_z$. FFT is done with 200-400 $T$.  One gets new peaks at $k{\omega}/{2}\pm\Omega$,  indicating the existence of time quasi-crystal. Such a new phase will not be destroyed given perturbations as long as they are small. Inset is the dynamics for $\epsilon_z=0.04, 0.11$. }
	\label{fig.perturb}
\end{figure}

	\subsection{Imperfect Rotation}

	Besides the perturbative interaction, we also consider the rotation perturbation and add $-h_2(t)\epsilon_z \sum_j \cos(\Omega t)\sigma_j^z$ to $H_{d}$ in Eq.~\ref{eq.H_d} with uniform strength $\epsilon_z$.
	The Fourier spectrum is shown in Fig.~\ref{fig.perturb} for various $\epsilon_z$, and the inset shows the quasi-periodically oscillating evolution of $S^y(t)$.  Similar as the previous case, subharmonic peaks are locked at  $k\omega/2\pm\Omega$ with descending heights for increasing perturbations. We find that for $\epsilon_z=0.04$, the oscillations persist without decay up to at least 1000$T$, while the rapid decay with $\epsilon_z=0.11$ suggests that the system heats up quickly.

\begin{figure}[h]
	\centering
	\includegraphics[width=0.8\linewidth]{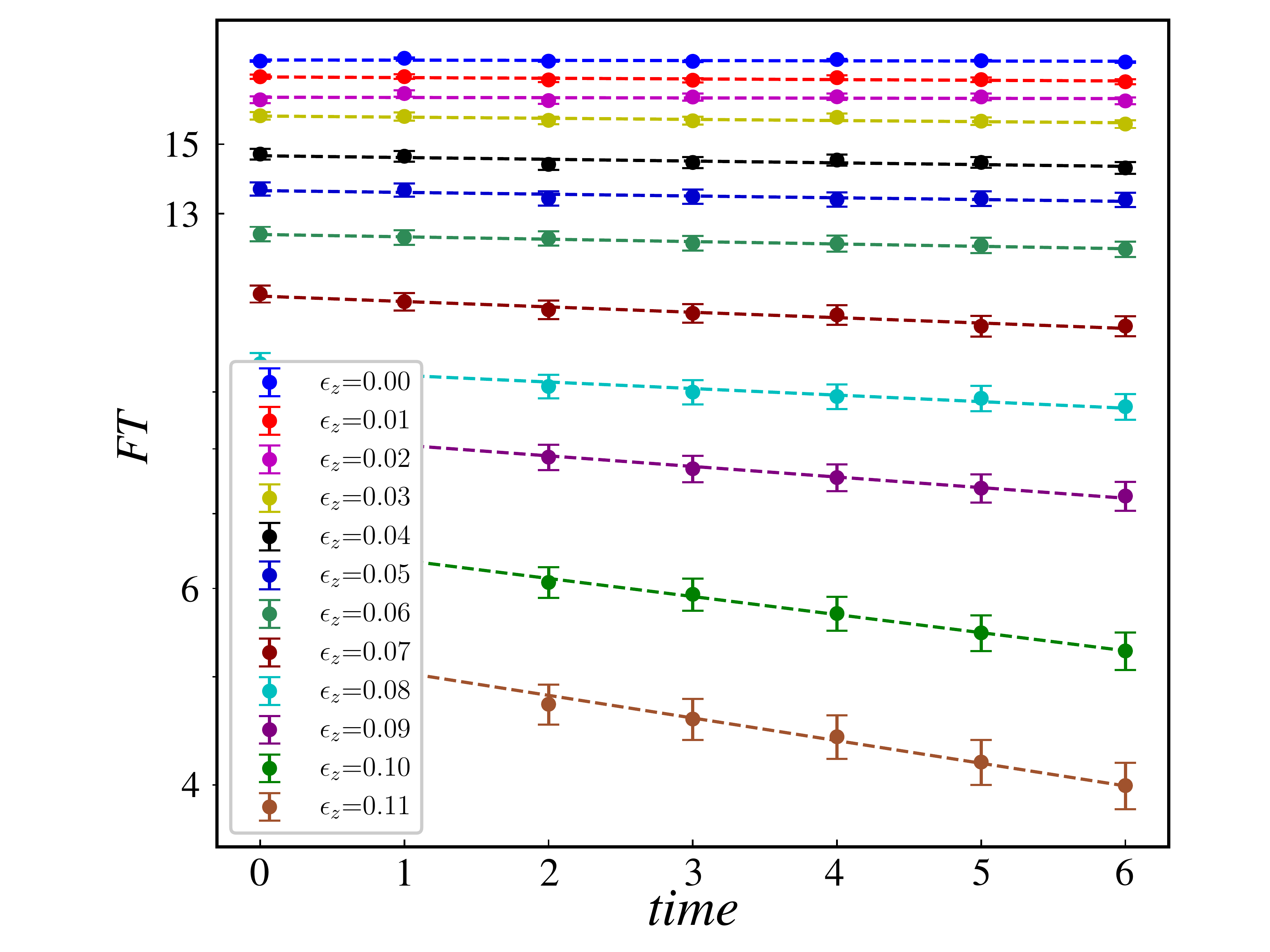}
	\caption{Amplitudes of the first subharmonic peak of the Fourier spectrum of $S_y$ for the DTQC versus time (300 disorder configurations) for different pulse perturbations.}
	\label{fig:fftpeak}
\end{figure}

\begin{figure}
	\centering
	\includegraphics[width=0.8\linewidth]{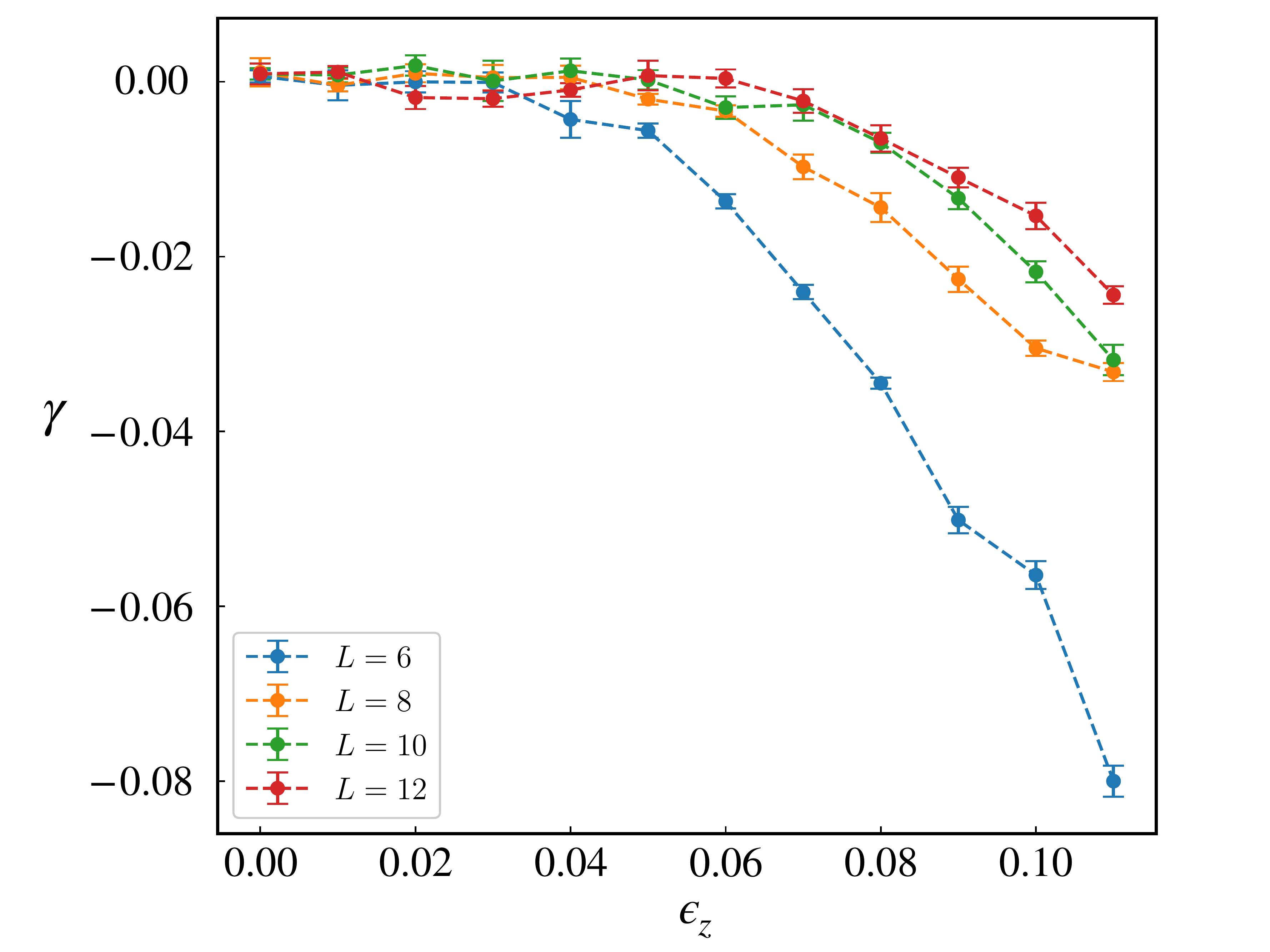}
	\caption{Decay rate $\gamma$ of the first subharmonic peak in Fourier spectrum versus perturbation $\epsilon_z$ for varying system sizes. $\gamma$ remains zero for for the DTQC and becomes negative for the thermalised phase, and the transition point locates around $\epsilon_z\sim0.05$.}
	\label{fig:decay}
\end{figure}

	Long time behaviour of the first subharmonic peak computed with 10 sites for various perturbations are plotted in Fig.~\ref{fig:fftpeak}. The same method to obtain average and error bar is applied here as above, and a regime with zero $\gamma$ can be observed. The relation between $\gamma$ and $\epsilon_z$ is depicted in Fig.~\ref{fig:decay}. The critical value of the thermal phase transition is approximately $\epsilon_z\sim0.05$.
\end{document}